# Specific Heat and Thermal Entanglement in an Open Quantum system

Behzad Lari[*]

Department of Physics, Ahvaz Branch, Islamic Azad University, Ahvaz, Iran

E-mail: behzadlari1979@yahoo.com [*]

## Abstract

In this brief report, we attention to the system of two qubits modeled by Heisenberg XXZ chain with the Dzyaloshinskii–Moriya interaction. The system exposed to bosonic baths with the Cauchy-Lorentz distribution of frequency. We've got a new formula to calculate the specific heat of open systems using density and found that the specific heat at low temperatures can be negative. We observed that when the state of system is entangle, in contradiction with the third law of thermodynamics, the specific heat is diverge when the temperature goes to zero. The speed of divergence is depend to the amount of entanglement. These results may be useful to design solid quantum gate and quantum memories.

**Keywords:** Open Quantum System, Dzyaloshinskii–Moriya Interaction, Born Markove approximation, Entanglement, Specific Heat.

## 1. Introduction

To design solid quantum gates and memories [1], [2], understanding behavior of the standard interaction models in solids and analyze the relation between amount entanglement and thermodynamics parameters of the system seem to be important. One of the standard models is anisotropic Heisenberg XXZ model for two qubits with the Dzyaloshinskii–Moriya (D-M) interaction parameter (arising from the spin–orbit coupling) [3]. In special, when each qubits of system exposed to separate bosonic bath. Many scientific reports have paid attention to different open quantum system and their amount of entanglement [4-6] but rare of them look for relation between thermal entanglement and specific heat as a determinative thermodynamics parameter of solids which is important to select material to use in manufacture of quantum gates (QG) and quantum memories (QM). Specially, in this article we established a method to calculate specific heat using density matrix of system after exposed to environment. To calculate the density matrix of principle system after evolution, we used a unitary transform to Interaction picture and solved the Von-Neumann equation in Born-Markov approximation. We have assumed that the bosonic baths with Cauchy-Lorentz distribution of frequency. Attention to evolution of entanglement with time and other effective parameters on Hamiltonian of the system as spin-orbit coupling $D_z$ and coupling constant $J_i$ (where $i \in \{x, y, z\}$) (which naturally depend to sample of solid materials) can help us in practical assembling of QG. This paper is organized as follows. In Sec. 2, we use the Hamiltonian of anisotropic Heisenberg XXZ model for two qubits with the D-M Interaction exposed to separate bosonic baths and we obtained initial density matrix of system as a function of temperature and we solved the Non-Markovian Master equation for our model. In Sec. 3, present a method to calculate specific heat using density matrix of open quantum system. Also, we calculate the entanglement of formation (EOF). Finally, we present our conclusions in Sec. 4.

## 2. Solution for anisotropic Heisenberg XXZ model with D-M interaction in bosonic baths via Non-Markovian Approach

The Hamiltonian of anisotropic Heisenberg XXZ model with D-M interaction in separate bosonic baths can introduce as follow,

$$H = H_s + H_b + H_{sb} \tag{1}$$

where
$$H_s = \sum_i J_i \, \sigma_1^i \sigma_2^i + D_z\big(\sigma_1^x \sigma_2^y - \sigma_1^y \sigma_2^x\big) \quad , i \in \{x,y,z\}) \tag{2}$$

$$H_b = \sum_{i=1}^{2} \omega_{n_i} b_{n_i}^\dagger b_{n_i} \tag{3}$$

$$H_{sb} = \sum_{i=1}^{2} \sigma_i^+ \sum_{n_i} g_{n_i} b_{n_i} + h.c. \tag{4}$$

Where $H_s$, describe the Hamiltonian of the system with D-M interaction and $H_b$ shows the Hamiltonian of baths with $n_i$ mode in frequency $\omega_{n_i}$. $H_{sb}$ denotes the Hamiltonian of interaction between system and baths with strength interaction $g_{n_i}$ where "i" is the label for first or second qubit and $\sigma_i$'s are the Pauli matrices. The coupling constant $J_i > 0$ correspond to the antiferromagnetic case and $J_i < 0$ correspond to the ferromagnetic case. When $J_x = J_y = J$ and $J_z \neq J$ the model is called Heisenberg XXZ model. The thermal density matrix of system in basis $\{|00\rangle, |10\rangle, |01\rangle, |11\rangle\}$ is given by,

$$\rho_s(t=0,T) = \frac{e^{-\beta H_s}}{Z} \tag{5}$$

$$= \frac{1}{Z}\{e^{-\beta J_z}|00\rangle\langle 00| + u|10\rangle\langle 10| + ve^{i\theta}|01\rangle\langle 10| + ve^{-i\theta}|10\rangle\langle 01| + u|01\rangle\langle 01| + e^{-\beta J_z}|11\rangle\langle 11|\}$$

Where $\beta = \frac{1}{k_B T}$ ($k_B$ is Boltzmann constant) and

$$u = \frac{1}{2} e^{\beta(J_z - 2\eta)}(1 + e^{4\beta\eta})$$

$$v = \frac{1}{2} e^{-\beta(J_z - 2\eta)}(1 - e^{4\beta\eta})$$

$$Z = 2\, e^{-\beta J_z}[1 + e^{2\beta J_z} Cosh(2\beta\eta)] \tag{6}$$

$$\eta = \sqrt{J^2 + D_z^2} \quad ; \quad \theta = tan^{-1}\left(\frac{D_z}{J}\right)$$

The density matrix in Eq. (5) has the X-type form [7]. We apply the initial density matrix to find its time evolution using the following Von-Liouville equation.

$$\frac{\partial \rho_{sb}(t)}{\partial t} = -\frac{i}{\hbar}[H(t), \rho_{sb}(t)] \tag{7}$$

In rest of this paper we set $\hbar = 1$. The Non-Markovian Master equation for density matrix of system in interaction picture with respect to Born-approximation is obtained as

$$\frac{\partial \rho_s^I(t)}{\partial t} = -i\, tr_b\{[H_{sb}^I(t), \rho_{sb}^I(0)]\} - \int_0^t d\acute{t}\, tr_b\{[H_{sb}^I(t), [H_{sb}^I(\acute{t}), \rho_{sb}^I(\acute{t})]]\} \tag{8}$$

The first term in Eq. (6) always is zero if the initial state select eigenstate. Finally

$$\frac{\partial \rho_s^I(t)}{\partial t} = -\int_0^t d\acute{t}\, tr_b\{[H_{sb}^I(t), [H_{sb}^I(\acute{t}), \rho_s^I(t) \otimes \rho_b^I(0)]]\} \tag{9}$$

where
$$H_{sb}^I(t) = \exp\{-i(H_s + H_b)t\}\, H_{sb}\, \exp\{i(H_s + H_b)t\} \tag{10}$$

$$\rho_s^I(t) = \exp\{-iH_s t\}\, \rho_{sb}(t)\, \exp\{iH_s t\} \tag{11}$$

$$\rho_b^I(0) = \rho_b(0) \tag{12}$$

The Eq. (4), can be rewritten in interaction picture,

$$H_{sb}^I = \sum_{j=1}^{2} \sigma_j^+ \sum_{n_j} g_{n_j} b_{n_j} \exp\{i(\varepsilon - \omega_{n_j})t\} + h.c. \tag{13}$$

We select the general initial state of the thermal bath as,

$$\rho_b(0) = |N; \ldots, n_j, \ldots, n_p, \ldots\rangle_b \langle N; \ldots, n_j, \ldots, n_p, \ldots| \tag{14}$$

Finally after trace out on baths part and using Eq. (14) we obtain

$$\frac{d\rho_s^I(t)}{dt} = \sum_{j=1}^{2}\{[\sigma_j^+, \rho_s^I(t)\sigma_j^-]\sum_{n_j}|g_{n_j}|^2 \int_0^t d\acute{t}\, e^{i(\varepsilon-\omega_{n_j})(t-\acute{t})} + h.c.\} = \sum_{j=1}^{2}\mathcal{J}_j(t)\rho_s^I(t) \tag{15}$$

Where $\mathcal{J}_j(t)$ are the super operators, and for a long time, are called the Lindblad operators.

If suppose the baths are bosonic with Cauchy-Lorentz distribution we have

$$\frac{d\rho_s^I(t)}{dt} = R(t)\sum_{j=1}^{2}[\sigma_j^+, \rho_s^I(t)\sigma_j^-] + h.c. = \sum_{j=1}^{2}\mathcal{J}_j(t)\rho_s^I(t)$$

where $\quad R(t) = \int_0^t d\acute{t}\int_{-\infty}^{+\infty} d\omega\, J(\omega)\, e^{i(\epsilon-\omega)\acute{t}} = \frac{\gamma_0}{2}(1 - e^{-\gamma t})$ \quad (16)

$\gamma$ is the scale parameter which specifies the half-width at half-maximum. Finally the following master equation is obtained in Schrödinger picture.

$$\frac{d\rho_s(t)}{dt} = -[H_s, \rho_s(t)] + \sum_{j=1}^{2}\mathcal{J}_j(t)\rho_s(t) \tag{17}$$

After a little algebra, we obtained three independent differential equation on components of $\rho_{ij}(t)$ as follows,

$$\frac{d}{dt}\begin{pmatrix}\rho_{14}(t,T)\\\rho_{41}(t,T)\end{pmatrix} = \begin{pmatrix}-\gamma_0(1-e^{-\gamma t}) & 0\\ 0 & -\gamma_0(1-e^{-\gamma t})\end{pmatrix}\begin{pmatrix}\rho_{14}(t,T)\\\rho_{41}(t,T)\end{pmatrix} \tag{18}$$

$$\frac{d}{dt}\begin{pmatrix}\rho_{12}(t,T)\\\rho_{13}(t,T)\\\rho_{24}(t,T)\\\rho_{34}(t,T)\end{pmatrix}=\begin{pmatrix}-2iJ_z-\frac{3\gamma_0}{2}(1-e^{-\gamma t}) & i(2J-2iD_z) & 0 & 0\\ i(2J+2iD_z) & -2iJ_z-\frac{3\gamma_0}{2}(1-e^{-\gamma t}) & 0 & 0\\ 0 & \gamma_0(1-e^{-\gamma t}) & 2iJ_z-\frac{\gamma_0}{2}(1-e^{-\gamma t}) & -i(2J+2iD_z)\\ \gamma_0(1-e^{-\gamma t}) & 0 & -i(2J-2iD_z) & -2iJ_z-\frac{\gamma_0}{2}(1-e^{-\gamma t})\end{pmatrix}\begin{pmatrix}\rho_{12}(t,T)\\\rho_{13}(t,T)\\\rho_{24}(t,T)\\\rho_{34}(t,T)\end{pmatrix} \quad (19)$$

$$\frac{d}{dt}\begin{pmatrix}\rho_{11}(t,T)\\\rho_{22}(t,T)\\\rho_{33}(t,T)\\\rho_{44}(t,T)\\\rho_{23}(t,T)\\\rho_{32}(t,T)\end{pmatrix}=\begin{pmatrix}-\frac{4\gamma_0}{2}(1-e^{-\gamma t}) & 0 & 0 & 0 & 0 & 0\\ \gamma_0(1-e^{-\gamma t}) & -\gamma_0(1-e^{-\gamma t}) & 0 & 0 & i(2J-2iD_z) & -i(2J+2iD_z)\\ \gamma_0(1-e^{-\gamma t}) & 0 & -\gamma_0(1-e^{-\gamma t}) & 0 & -i(2J-2iD_z) & i(2J-2iD_z)\\ 0 & \gamma_0(1-e^{-\gamma t}) & \gamma_0(1-e^{-\gamma t}) & 0 & 0 & 0\\ 0 & i(2J+2iD_z) & -i(2J+2iD_z) & 0 & -\gamma_0(1-e^{-\gamma t}) & 0\\ 0 & -i(2J-2iD_z) & i(2J-2iD_z) & 0 & 0 & -\gamma_0(1-e^{-\gamma t})\end{pmatrix}\begin{pmatrix}\rho_{11}(t,T)\\\rho_{22}(t,T)\\\rho_{33}(t,T)\\\rho_{44}(t,T)\\\rho_{23}(t,T)\\\rho_{32}(t,T)\end{pmatrix} \quad (20)$$

We can reduce the Eqs. (18), (19) and (20) to the following form but by attention to initial state on Eq. (5), only the elements of density matrix in last equation remain non zero.

$$\frac{d}{dt}|\alpha(t)\rangle_i = M_i(t)|\alpha(t)\rangle_i \quad (21)$$

Using the similarity transformation "P", we have

$$e^{M_i(t)} = P\,e^{G}P^{-1} \quad (22)$$

Where "G" is the Jordan form of $M_i(t)$. Finally we obtained the following relation to calculate non-zero elements of density matrix.

$$\begin{pmatrix}\rho_{11}(t,T)\\\rho_{22}(t,T)\\\rho_{33}(t,T)\\\rho_{44}(t,T)\\\rho_{23}(t,T)\\\rho_{32}(t,T)\end{pmatrix}=\begin{pmatrix}e^{-4B} & 0 & 0 & 0 & 0 & 0\\ -e^{-4B}+e^{-2B} & \frac{e^{-2B}}{2}(1+\cos[|\Gamma|.t]) & \frac{e^{-2B}}{2}(1-\cos[|\Gamma|.t]) & 0 & \frac{-e^{-2B}}{2}\frac{|\Gamma|}{4\Gamma}\sin[|\Gamma|.t] & \frac{-e^{-2B}}{2}\frac{|\Gamma|}{4\Gamma}\sin[|\Gamma|.t]\\ -e^{-4B}+e^{-2B} & \frac{e^{-2B}}{2}(1-\cos[|\Gamma|.t]) & \frac{e^{-2B}}{2}(1+\cos[|\Gamma|.t]) & 0 & \frac{e^{-2B}}{2}\frac{|\Gamma|}{4\Gamma}\sin[|\Gamma|.t] & \frac{e^{-2B}}{2}\frac{|\Gamma|}{4\Gamma}\sin[|\Gamma|.t]\\ 1+e^{-4B}-2e^{-2B} & 1-e^{-2B} & 1-e^{-4B} & 1 & 0 & 0\\ 0 & \frac{e^{-2B}}{2}\frac{|\Gamma|}{4\Gamma^*}\sin[|\Gamma|.t] & \frac{-e^{-2B}}{2}\frac{|\Gamma|}{4\Gamma}\sin[|\Gamma|.t] & 0 & \frac{e^{-2B}}{2}(1+\cos[|\Gamma|.t]) & \frac{-\Gamma^*}{\Gamma}\frac{e^{-2B}}{2}(1+\cos[|\Gamma|.t]))\\ 0 & \frac{e^{-2B}}{2}\frac{|\Gamma|}{4\Gamma}\sin[|\Gamma|.t] & \frac{-e^{-2B}}{2}\frac{|\Gamma|}{4\Gamma}\sin[|\Gamma|.t] & 0 & \frac{-\Gamma^*}{\Gamma}\frac{e^{-2B}}{2}(1+\cos[|\Gamma|.t]) & \frac{e^{-2B}}{2}(1+\cos[|\Gamma|.t])\end{pmatrix}\begin{pmatrix}\rho_{11}(0,T)\\\rho_{22}(0,T)\\\rho_{33}(0,T)\\\rho_{44}(0,T)\\\rho_{23}(0,T)\\\rho_{32}(0,T)\end{pmatrix}$$
(23)

Where
$$B = \frac{\gamma_0}{\gamma}t + \frac{\gamma_0}{2\gamma}(e^{-\gamma t}-1)$$

$$\Gamma = i\,\eta$$

$$|\Gamma| = \sqrt{\Gamma\Gamma^*}$$

Where the $\rho_{ij}(0,T)$ are components of density matrix introduced on Eq.(5).

### 3. specific heat of an open quantum system

In this section we proposed a method to calculate the specific heat of open quantum system using its density matrix. We used the following partition function [8],

$$Z_S = \frac{Tr_{sb}[\exp(-\beta H)]}{Tr_b[\exp(-\beta H_B)]} \quad (24)$$

Where the H defined in Eq. (1) and $H_b$ is Hamiltonian of the baths. The specific heat is defined as

$$C = k_B \beta^2 \frac{\partial^2}{\partial \beta^2} Ln(Z) \qquad (25)$$

The specific heat of principle system is difference between the specific heat of total system and baths.

$$C_s(t,T) = C_{sb}(t,T) - C_b(t,T) \qquad (26)$$

Since, the general thermal density matrix at time "t" is obtained as,

$$\rho(t,T) = \frac{1}{Z(t,T)} \exp[-\beta H] \qquad (27)$$

Using the Eqs. (25), (26) and (27) we have

$$C_s(t,T) = -4 \, k_B \, \beta^2 \, \frac{\partial^2}{\partial \beta^2} ( Tr[\, Ln(\rho_{sb})\,] - Tr_b[\, Ln(\rho_b)\,]\,). \qquad (28)$$

When the interaction between system and baths is weak, we obtained

$$C_s(t,T) = -4 \, n \, k_B \, \beta^2 \, \frac{\partial^2}{\partial \beta^2} \sum_{i=1}^{m} Ln(\eta_i). \qquad (29)$$

Where m (=4) is dimension of Hilbert space of the system. $\eta_i's$ are eigenvalues of density matrix of the system. The "n" is total number of modes in bath, $k_B$ is Boltzmann constant and $\beta = \frac{1}{k_B T}$.

We introduce $C_s^{(n)}(t,T) = \frac{C_s(t,T)}{4nk_B}$ and plot it as a function of time and $k_B T$.

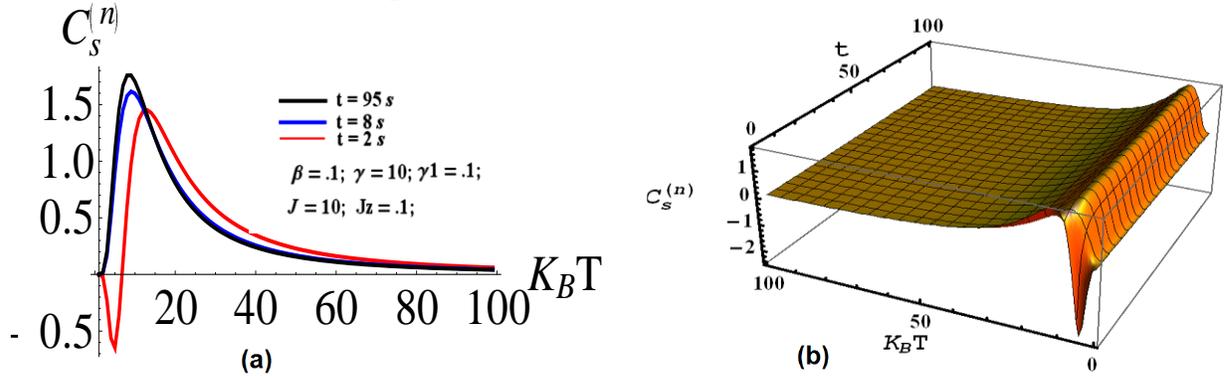

Figure 1. (a) $C_s^{(n)}$ versus $k_B T$ in different time (b) $C_s^{(n)}$ in time and $k_B T$. As seen in low temperatures, the specific heat can be negative. This is because the system has been exposed by environment.

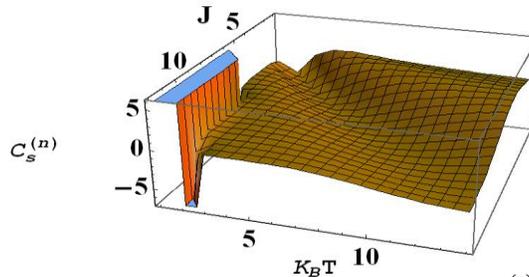

Figure 2. $C_s^{(n)}$ as a function of temperature and J. As we can see for J >7 and in low temperature the $C_s^{(n)}$ is diverged. This in contradiction with the third law of thermodynamics and is related to the existence of entanglement in system.

In order to quantify entanglement in D-M model which was solved in section II, we use the EOF derived by [9], [10]

$$EN = h\left(\frac{1+\sqrt{1-C^2}}{2}\right) \qquad (30)$$

Where $h(x) = -x\,Log_2 x - (1-x)\,Log_2(1-x)$ and $C = \max\{0, \lambda_1 - \lambda_2 - \lambda_3 - \lambda_4\}$ is the concurrence and $\lambda_i (i=1,2,3,4)$, are the square roots of the eigenvalues of the operator $\rho_s(t,T)\widetilde{\rho_s(t,T)}$ in descending order with $\lambda_1 \geq \lambda_2 \geq \lambda_3 \geq \lambda_4$. The $\tilde{\rho}_s(t,T)$ defined as

$$\tilde{\rho}_s(t,T) = (\sigma_{1y} \otimes \sigma_{2y})\, \rho_s^*(t,T)\, (\sigma_{1y} \otimes \sigma_{2y})$$

$\rho(t,T)$ is the density matrix of system which obtained in D-M model. The density matrix of the system has the X- type form [7]. $\sigma_{1y}$ and $\sigma_{2y}$ are the normal Pauli operators. We plotted the EOF in following figures.

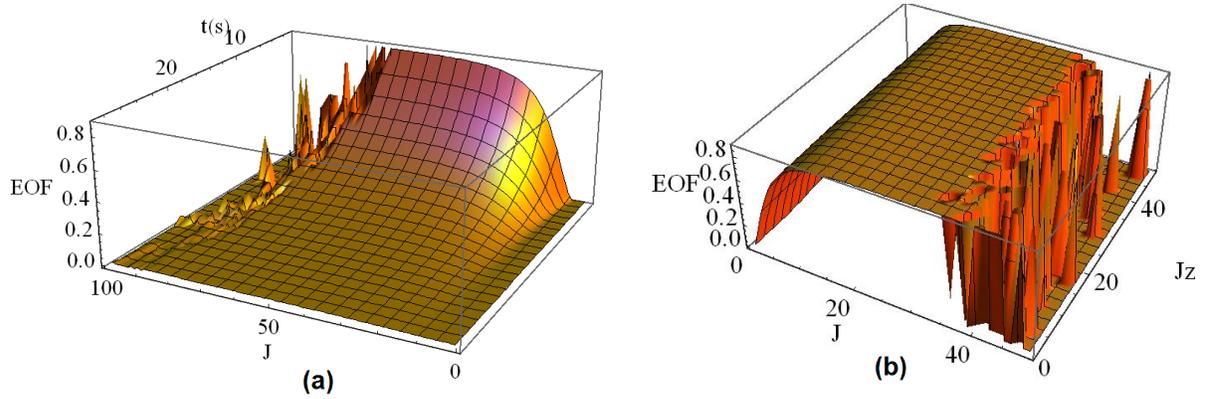

Figure 3. The plot of EOF at $\beta = 0.1$ as a function of (a) Coupling constant J and time (b) Coupling constant J and $J_z$. As seen the entanglement of formation is maximum for all $J_z$ and large amount of J (for J>11).

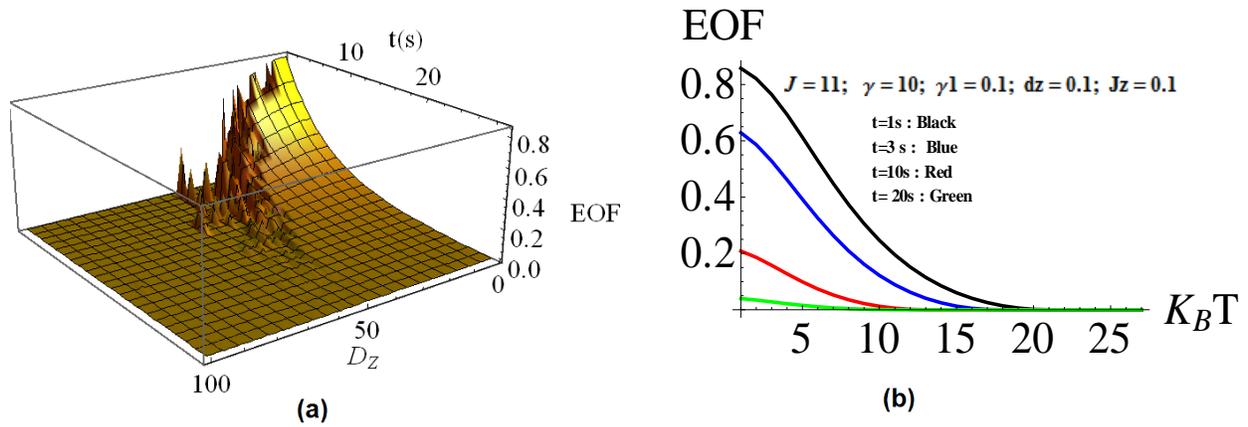

Figure 4. (a) The plot of EOF (for $\beta = 0.1$) as a function of Spin-Orbit coupling $D_z$ and time. In small $D_z$, the EOF is maximum and decrease with time monotonically. (b) The plot of EOF as a function of temperature in different time. We can see in zero temperature, the entanglement is maximum.

## 4. Results and Conclusions

According to our results, when the system with D-M interaction exposed to environment, the amount of EOF on system decrease with time in different J. In big J, the EOF successively disappeared which is not good report to design the stable quantum gates and quantum memories. As we can see on Fig. (3b), the EOF is not sensitive to variation of $J_z$. It seems that on small $D_z$ and low temperature, the EOF is maximum and decreased with time monotonically. In small temperature, when the EOF to be max (Fig.4b), the specific heat in conflict with the third law of thermodynamics is diverged (Fig.2). This divergence is due to entanglement in the system for closed systems has also been reported [11]. We observed negative specific heat that is because the system with D-M regime has been exposed to environment [8] (Fig.1). According to the Figs (2) and (3a) we can see that for bigger "J" the amount of entanglement is more and thus the speed of divergence in the specific heat near zero temperature increases. All these results may have a better understanding of the system designed to provide stable quantum gates.

**ACKNOWLEDGMENTS**

We gratefully acknowledge fruitful discussions with Prof. H. Hassanabadi and K. Zaheeri.